\documentclass[10pt]{article}
\usepackage{amsmath}
\usepackage{amssymb}
\usepackage{graphics}
\usepackage{rotating}
\usepackage{cite}
\usepackage{color}
\usepackage{fancybox}
\usepackage{pstricks}
\usepackage{color, colortbl}

\usepackage{indentfirst}

\definecolor{ay}{rgb}{0.91, 0.84, 0.42}
\definecolor{arsenic}{rgb}{0.23, 0.27, 0.29}
\definecolor{darkseagreen}{rgb}{0.21, 0.37, 0.23}
\definecolor{darkred}{rgb}{0.55, 0.0, 0.0}
\definecolor{asparagus}{rgb}{0.53, 0.66, 0.42}
\definecolor{med}{rgb}{0.79, 0.86, 0.54}
\definecolor{arylideyellow}{rgb}{0.91, 0.84, 0.42}

\def\0{\mbox{\tiny $0$}}
\def\1{\mbox{\tiny $1$}}
\def\2{\mbox{\tiny $2$}}
\def\3{\mbox{\tiny $3$}}
\def\4{\mbox{\tiny $4$}}
\def\5{\mbox{\tiny $5$}}
\def\6{\mbox{\tiny $6$}}
\def\7{\mbox{\tiny $7$}}
\def\8{\mbox{\tiny $8$}}
\def\9{\mbox{\tiny $9$}}
\title{\hspace*{-.5cm}\shadowbox{\fcolorbox{black}{ay} { {\color{arsenic}{ \large \textbf{
\begin{tabular}{c}
THE EFFECT OF THE GEOMETRICAL OPTICAL PHASE\\ ON THE PROPAGATION  OF HERMITE-GAUSSIAN BEAMS\\ THROUGH TRANSVERSAL AND PARALLEL DIELECTRIC BLOCKS
\end{tabular}}}}}}
}
\author{
\small Silv\^ania A. Carvalho$^{\1}$\,\,\,\,and\,\,\,\,Stefano De Leo$^{\2}$ \\
\small $^{\1}$ Department of Exact Sciences, Fluminense Federal University, Volta Redonda (Brazil)\\
\small $^{\2}$ Department of Applied Mathematics, Campinas State University , Campinas (Brazil)}
\date{\small
\fcolorbox{black}{med} {\color{darkred} $\bullet$ {\color{arsenic}{
{\small \textbf{Journal of Modern Optics 66, 548-556 (2019)}}}} {\color{darkred}{ $\bullet$}} } }

\begin{document}


\maketitle

\vspace*{-.7cm}

\begin{abstract}
\noindent {When an optical beam propagates through dielectric blocks, its optical phase is responsible for  the path of the beam. In particular, the first order Taylor expansion of the geometrical part reproduces the path predicted by the Snell and reflection laws whereas the first order expansion of the Fresnel phase leads to the Goos-H\"anchen shift. In this paper, we analyze the effects of the second order Taylor expansion of the geometrical phase  on the shape of the optical beam and  show how it affects the transversal symmetry of Hermite-Gaussian beams. From the analytical expression of the transmitted beam, it is possible to determine in which transversal and parallel dielectric blocks configuration the transversal symmetry breaking is maximized or when the symmetry is recovered. We also discuss the axial spreading delay.}

\end{abstract}






\section*{\color{darkred} \normalsize I. INTRODUCTION}

Over the years, studies on  Gaussian beams propagating through dielectric blocks have been the subject matter of great interest due to their applications in Optics\cite{Snell1948,Snell1951,Snell1958,Snell1980,Snell2005,Snell2007}, in particular in the study of deviations from the Snell and reflection laws\cite{AJPBlock,2013JMO,2014JO16,2014PRA90}. In \cite{AJPBlock}, the stationary phase method was applied to the geometrical and Fresnel phases showing how the first order Taylor expansion leads to the optical path predicted by the geometrical optics and how, in the case of total reflection, the first order Taylor expansion of the Fresnel coefficients determines the additional lateral displacement, known in literature as the Goos-H\"anchen  shift. Recent studies\cite{2016JMOTSB,2017JMOETSB} investigated  the effects of the second order term of the geometrical phase for the propagation of Gaussian beams through  transversal dielectric blocks.

In this article, we extend  the study of the second order term of the optical geometrical phase to Hermite-Gaussian beams propagating through  a mixed configuration of  transversal  and parallel dielectric blocks. These beams play an important role in the development of resonators\cite{Saleh2007,Wolf1999,HGLG2005,HGLG2009}, as well as in the production of  Laguerre-Gaussian beams \cite{Saleh2007,HGLG2005,HGLG2009,HGLG1998}. So, a correct description of their propagation through dielectric blocks and an understanding on how their shape is modified by a mixed transversal and parallel configuration of dielectric blocks is surely important in view of their applications. Although several articles discuss  Hermite-Gaussian beams in some detail\cite{Saleh2007,Wolf1999,HGLG2004}, there are certain aspects such as a closed analytical formula for the  beam parameter control that surely deserves more attention. This contribution aims to cover some of these aspects in detail.

In section II, we introduce the analytical formula used to describe the propagation
of Hermite-Gaussian beams in  free space. In section III, based on the analogy between Optics and Quantum Mechanics, we give the Fresnel coefficients for transverse electric (TE) and transverse magnetic (TM) waves and introduce the optical geometrical phase determined by the geometrical properties of the blocks, in particular by their air/dielectric and dielectric/air interface positions
\cite{AJPBlock,2013JMO,2014JO16,2014PRA90,2016JMOTSB,2011EPJD61,2013EPJD67}.
This approach uses the Maxwell equations for photon propagation in the presence of a dielectric block to mimic the quantum-mechanical Schr\"odinger equation describing the electron propagation in the presence of a potential step. In this formulation, the transmission coefficient is determined not only by the Fresnel coefficients but also by the geometrical phase coming from the continuity condition at each interface. The Taylor  expansion of this phase and the Fresnel coefficients  allow  an analytical expression for the intensity of the transmitted Hermite-Gaussian beam\cite{JOSA67a1977,HG2012,PJP2011,HG2013}. In section IV, we introduce the angular notation to calculate the first and second order contribution of the optical phase.
As observed before, the first order contribution of the optical phase is responsible, in its geometrical part, for the optical path \cite{Saleh2007} predicted by the Snell and reflection laws and, in its Fresnel part,  for the additional lateral displacement known as Goos-H\"anchen shift \cite{GHS1947,1948AP437,GHS1988,GHS2012,GHS2013,2013JMO,2014JO16}. The second order contribution acts on the transversal symmetry and on the  axial spreading  of the optical beam\cite{JOSA67a1977}. In section V, once obtained the analytical formula for the transmitted Hermite-Gaussian beam propagating through a mixed configuration of transversal and parallel dielectric blocks, we discuss our results and show how to use block rotations to control the spreading factor. This is done by simulating optical experiments with blocks of borosilicate glass (BK7),  He-Ne laser with $\lambda=633$\,{\rm nm}) and ${\rm w}_{\0}=200\,\mu{\rm m}$. Conclusions and outlooks are presented in the final section.

\section*{\color{darkred} \normalsize II. FREE PROPAGATION OF HERMITE-GAUSSIAN BEAMS}

The Gaussian electric field, representing the lower order of Hermite-Gaussian beams, may be generalized to higher modes by including an additional modulation function to the Gaussian wave number distribution
\begin{eqnarray}
\label{eq:gd}
G\left(k_{x},k_{y};\boldsymbol{r}\right) &=& \, \exp\left[\,-\,\frac{\mbox{w}_{\0}^{^{2}}}{4}\,\left(k_{x}^{^2} + k_{y}^{^2}\right)\,+ \,i\, \,\boldsymbol{k}\,\cdot\,\boldsymbol{r}\right]\,\,,
\end{eqnarray}
where $\mbox{w}_{\0}$ is the radius of the  beam waist and $|\boldsymbol{k}|=2\,\pi/\lambda$ with $\lambda$ the wavelength of the beam.  As shown in \cite{HGLG2004}, the wave number distribution generating all the modes of Hermite-Gaussian beams is given by
\begin{equation}
\label{mod}
\mathcal{H}_{_{\ell m}}\left(k_{x},k_{y};\boldsymbol{r}\right) = \left(i\,k_{x}\,\mbox{w}_{\0} + \displaystyle{\frac{i}{\mbox{w}_{\0}} \frac{\partial}{\partial k_{x}}}\,\right)^{\ell} \left(i\,k_{y}\,\mbox{w}_{\0} + \displaystyle{\frac{i}{\mbox{w}_{\0}} \frac{\partial}{\partial k_{y}}}\,\right)^{m} \,\, G\left(k_{x},k_{y};\boldsymbol{r}\right)\,\,,
\end{equation}
 where the indexes $\ell,m$ are associated with the Hermite-Gaussian modes.

The electric Hermite-Gaussian field is then obtained by integrating the modulated wave number distribution\cite{HGLG2005,HGLG2009}
\begin{equation}
\label{eq:HparaxAp}
E_{_{\ell m}} (\boldsymbol{r}) = N_{_{\ell m}} E_{\0}\, \displaystyle{\frac{\mbox{w}_{\0}^{^{2}}}{4\,\pi}} \int^{^{+\infty}}_{_{-\infty}}\hspace*{-.5cm}\mbox{d}k_x \int^{^{+\infty}}_{_{-\infty}}\hspace*{-.5cm} \mbox{d}k_y \, \left(i\,k_{x}\,\mbox{w}_{\0} + \displaystyle{\frac{i}{\mbox{w}_{\0}} \frac{\partial}{\partial k_{x}}}\,\right)^{\ell} \left(i\,k_{y}\,\mbox{w}_{\0} + \displaystyle{\frac{i}{\mbox{w}_{\0}} \frac{\partial}{\partial k_{y}}}\,\right)^{m} \hspace*{-.05cm}\,G\left(k_{x},k_{y};\boldsymbol{r}\right)\,\,,
\end{equation}
where $N_{_{\ell m}}$ is the normalization constant. In the paraxial approximation, $k_z\approx k - (k_x^{^{2}}+k_y^{^{2}})/2\,k$, we have\cite{HGLG2004}
\begin{equation}
\left[i\,k_{x}\,\mbox{w}_{\0} + \displaystyle{\frac{i}{\mbox{w}_{\0}} \frac{\partial}{\partial k_{x}}}\right] \,G\left(k_{x},k_{y};\boldsymbol{r}\right) = \displaystyle{\left[i\,k_{x}\,\mbox{w}_{\0}\,f(z) - \frac{x}{\mbox{w}_{\0}} \right]} \,G\left(k_{x},k_{y};\boldsymbol{r}\right)\,\,,
\label{eq:Iden}
\end{equation}
where
\begin{equation}
\label{eq:f}
2\,f(z) = 1 - 2\,i\,z / k\,\mbox{w}_{\0}^{^{2}} = 1 - i\,z / z_{_{R}}\,\,.
\end{equation}
By observing that $i\, k_{x}$ in the identity (\ref{eq:Iden})  can be substituted by
$\partial / \partial x$ , we can rewrite Eq.(\ref{eq:HparaxAp}) as follows
\begin{equation}
\label{eq:HparaxSimp}
E_{_{\ell m}} (\boldsymbol{r}) = N_{_{\ell m}}\, \left[\,\mbox{w}_{\0}\,f(z)\,\right]^{^{\ell + m}}\, \left[\displaystyle{\frac{\partial}{\partial x} - \frac{x}{\mbox{w}_{\0}^{^{2}}\,f(z)}}\,\right]^{\ell}\,\left[\displaystyle{\frac{\partial}{\partial y} - \frac{y}{\mbox{w}_{\0}^{^{2}}\,f(z)}}\,\right]^{m}\,E(\boldsymbol{r})\,\,,
\end{equation}
where
\begin{equation}
E(\boldsymbol{r}) =  E_{\0}\, \displaystyle{\frac{\mbox{w}_{\0}^{^{2}}}{4\,\pi}} \int^{^{+\infty}}_{_{-\infty}}\hspace*{-.5cm}\mbox{d}k_x \int^{^{+\infty}}_{_{-\infty}}\hspace*{-.5cm} \mbox{d}k_y \, \hspace*{-.05cm}\,G\left(k_{x},k_{y};\boldsymbol{r}\right) = \displaystyle{\frac{E_{\0}\,\,e^{ikz}}{1 + \,i\,z / z_{_{R}}}} \,\exp\left[\,-\displaystyle{\frac{x^{^{2}}+\,\,y^{^{2}}}{\mbox{w}_{\0}^{^{2}} \left(1 + \,i\,z / z_{_{R}}\right)}}\,\right]\,\,.
\end{equation}
By using the operator identity
\begin{equation}
\label{eq:oper}
\left(\displaystyle{\frac{\partial}{\partial x} - 2\,\alpha\,x}\,\right)^{r} = \left(e^{\alpha \, x^{2}} \,\frac{\partial}{\partial x}\, e^{-\,\alpha \, x^{2}}\right)^{r} = \,e^{\alpha \, x^{2}} \,\frac{\partial^{^{r}}}{\partial x^{^{r}}}\, e^{-\,\alpha \, x^{2}}\,\,,
\end{equation}
we obtain
\begin{eqnarray}
\label{eq:HSimp2}
\hspace*{-.45cm}E_{_{\ell m}} (\boldsymbol{r}) &\hspace*{-.2cm}=\hspace*{-.2cm}& N_{_{\ell m}}\,\left[\,\mbox{w}_{\0}\,f(z)\,\right]^{^{\ell + m}}\, \exp\left[\,\displaystyle{\frac{x^{^{2}}+\,\,y^{^{2}}}{\mbox{w}_{\0}^{^{2}}\,\left(1 - i\,z / z_{_{R}}\right)}}\,\right]\,\frac{\partial^{^{\ell}}}{\partial x^{^{\ell}}} \,\frac{\partial^{^{m}}}{\partial y^{^{m}}}\,\exp\left[\,- \displaystyle{\frac{x^{^{2}}+\,\,y^{^{2}}}{\mbox{w}_{\0}^{^{2}}\,\left(1 - i\,z / z_{_{R}}\right)}}\,\right]\,E(\boldsymbol{r})\,\,.
\end{eqnarray}
\noindent The above expression can be further simplified by using the relation
\begin{equation}
\label{eq:rel1}
\displaystyle{\frac{x^{^{2}}+\,\,y^{^{2}}}{\mbox{w}_{\0}^{^{2}} \left(1 - \,i\,z / z_{_{R}}\right)}} + \displaystyle{\frac{x^{^{2}}+\,\,y^{^{2}}}{\mbox{w}_{\0}^{^{2}} \left(1 + \,i\,z / z_{_{R}}\right)}}\,=\,\displaystyle{\frac{2\,(x^{^{2}}+\,\,y^{^{2}})}{\mbox{w}^{^{2}}(z)}}\,\,,
\end{equation}
where
\[
\mbox{w}(z) = \mbox{w}_{\0}\, \sqrt{1\,+\,\left(\displaystyle{\frac{z}{z_{_{R}}}}\right)^{^{2}}}\,\,.
\]
The Hermite-Gaussian electric field is thus represented by
\begin{eqnarray}
\label{eq:HSimp3}
E_{_{\ell m}} (\boldsymbol{r}) &\hspace{-0.25cm}=\hspace{-0.25cm}& \,N_{_{\ell m}}\,\left[\,\mbox{w}_{\0}\,f(z)\,\right]^{^{\ell + m}}\, E(\boldsymbol{r})\,
\exp\left[\,\displaystyle{\frac{2\,(x^{^{2}}+\,\,y^{^{2}})}{\mbox{w}^{^{2}}(z)}}\,\right]\,\frac{\partial^{^{\ell}}}{\partial x^{^{\ell}}} \,\frac{\partial^{^{m}}}{\partial y^{^{m}}}\,\exp\left[\,- \displaystyle{\frac{2\,(x^{^{2}}+\,\,y^{^{2}})}{\mbox{w}^{^{2}}(z)}}\,\right]\,\,.
\end{eqnarray}
Finally by observing that
\begin{equation}
\label{eq:operH}
H_{_{m}}(u) = (-1)^{^{m}}\,e^{u^{2}} \,\frac{\partial^{^{m}}}{\partial u^{^{m}}}\, e^{-u^{2}}\,\,,
\end{equation}
we obtain
\begin{eqnarray}
\label{eq:HSimp5}
E_{_{\ell m}} (\boldsymbol{r}) &\hspace{-0.25cm}=\hspace{-0.25cm}& N_{_{\ell m}}\,E_{\0}\, \displaystyle{\frac{\mbox{w}_{\0}}{\mbox{w}(z)}} \,\displaystyle{\frac{(-1)^{^{\ell+m}}}{2^{^{(\ell+m)/2}}}}\,H_{_{\ell}}\left[\displaystyle{\frac{\sqrt{2}\,x}{\mbox{w}(z)}}\right]\,H_{_{m}}\left[\displaystyle{\frac{\sqrt{2}\,y}{\mbox{w}(z)}}\right] \, \exp\left[\,- \displaystyle{\frac{x^{^{2}}+y^{^{2}}}{\mbox{w}^{^{2}}(z)}}\,\right] \, \times \nonumber \\ &\hspace{-0.25cm}\hspace{-0.25cm}& \times \, \exp\left[\,i\, k\,z + i\,k\,\displaystyle{\frac{x^{^{2}}+y^{^{2}}}{2\,R(z)}} - i\,\left(\ell + m + 1 \right)\,\zeta(z)\,\right]\,\,,
\end{eqnarray}
 where $R(z) = z \left(1 + z_{_{R}}^{^{2}}/z^{^{2}} \right)$ is the radius of curvature of the wavefront and  $\zeta = \arctan(z/z_{_{R}})$ the Gouy phase\cite{Saleh2007}. The normalization constant is fixed by the power condition
\begin{equation}
\label{eq:normal}
P = \int^{^{+\infty}}_{_{-\infty}} \hspace{-0.5cm} \mbox{d}x  \int^{^{+\infty}}_{_{-\infty}} \hspace{-0.5cm} \mbox{d}y \,\left|E_{_{\ell m}}(\boldsymbol{r})\right|^{\2} = \frac{\pi \, \mbox{w}_{\0}^{\2}\,\left|E_{\0}\right|^{^{2}}}{2} \,\,.
\end{equation}
After simple algebraic manipulation, by using
\[
\, \int^{^{+\infty}}_{_{-\infty}} \hspace{-0.25cm} H_{_{p}}(x)\,H_{_{r}}(x) \,e^{-x^{2}}\,\mbox{d}x  = \sqrt{\pi}\, 2^{^{r}}\,r!\,\delta_{_{p r}} \,\,,
\]
we obtain $N_{_{\ell m}} = 1/\sqrt{m!\,\ell !}$.

The intensity of the Hermite-Gaussian electric field is then given by
\begin{eqnarray}
\label{eq:IHlm}
I_{_{\ell m}} (\boldsymbol{r}) &\hspace{-0.25cm}=\hspace{-0.25cm}& \,I_{\0}\,\displaystyle{\left[\frac{\mbox{w}_{\0}}{\mbox{w}(z)}\right]^{^{2}}} \,\displaystyle{\,\frac{1}{2^{^{\ell+m}}\,\ell!\,m!}}\,H_{_{\ell}}^{^{2}}\left[\displaystyle{\frac{\sqrt{2}\,x}{\mbox{w}(z)}}\right]\,H_{_{m}}^{^{2}}\left[\displaystyle{\frac{\sqrt{2}\,y}{\mbox{w}(z)}}\right]\, \exp\left[\,-\, 2 \,\displaystyle{\frac{x^{^{2}}+\,\,y^{^{2}}}{\mbox{w}^{^{2}}(z)}}\,\right]\,\,.
\end{eqnarray}

\section*{\color{darkred} \normalsize III. FRESNEL COEFFICIENTS AND OPTICAL PHASE}

\begin{figure}
\includegraphics[scale=0.72]{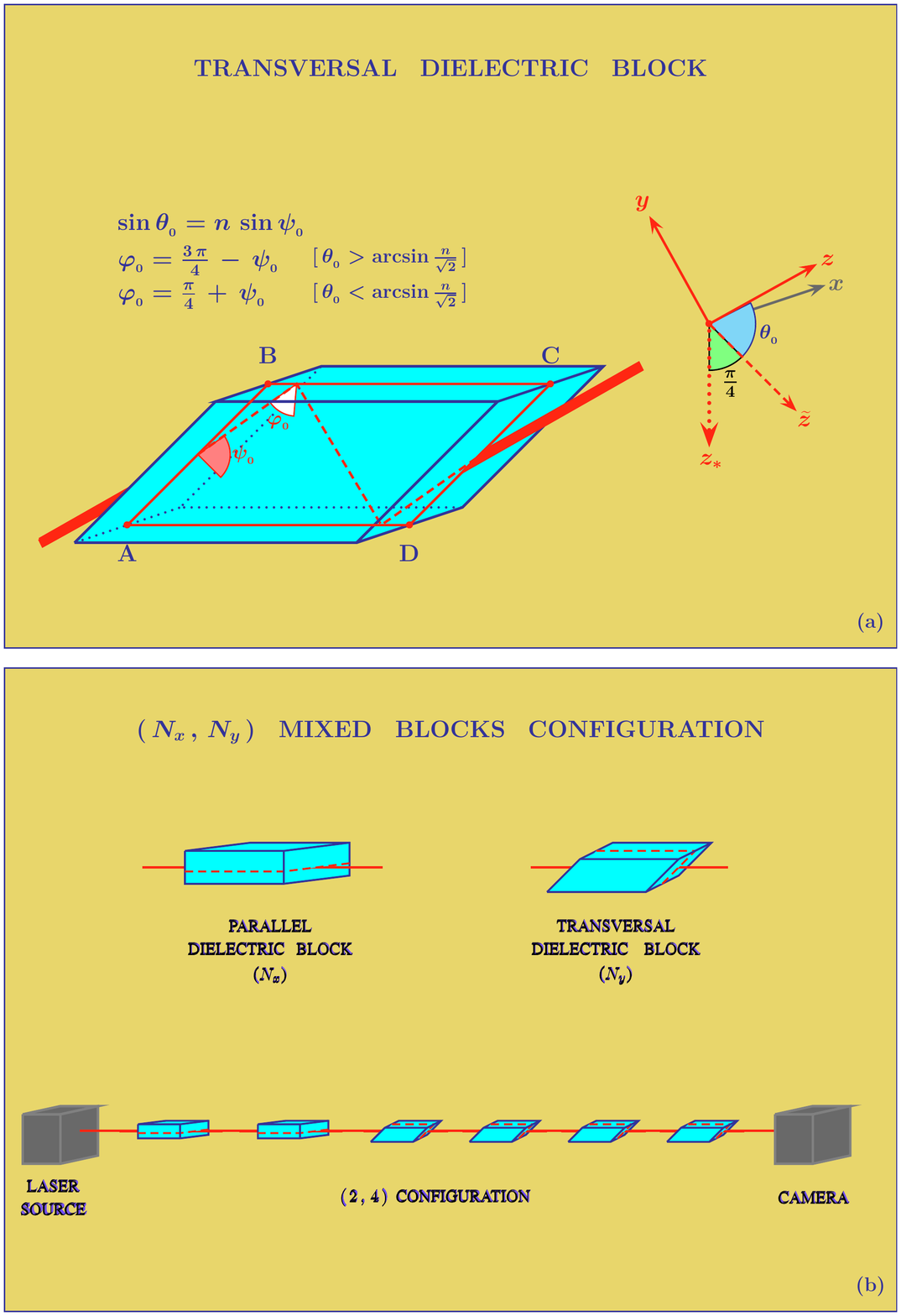}
\caption{\textbf{Geometrical layout of transversal and parallel dielectric blocks.} In (a), we plot the transversal dielectric block with its $yz$ plane of incidence. In (b), we show the $(N_{_{x}}\,,\,N_{_{y}})$ mixed, parallel and transversal, blocks configuration.}
\label{fig:Fig0}
\end{figure}

Before discussing the propagation of optical Hermite-Gaussian beams through a sequence of transversal and parallel dielectric blocks, let us introduce our elementary block \cite{2016JMOTSB,2017JMOETSB}. This block is a $45$ degree prism, see Fig.\,\ref{fig:Fig0}(a), built to guarantee two internal reflections. This is done by imposing the following geometrical constraint between the sides $\overline{AB}$ and $\overline{BC}$ of the block\cite{2016JMOTSB,2017JMOETSB}
\begin{equation}
\overline{BC} = \sqrt{2}\,\tan \varphi_{\0} \, \overline{AB} \,\,,
\label{eq:eqAD}
\end{equation}
where $\varphi_{\0}=\frac{3\,\pi}{4}-\psi_{\0}$ ($\theta_{\0}>\arcsin\frac{n}{\sqrt{2}}$) is the angle of incidence at the up and $\varphi_{\0}=\frac{\pi}{4}+\psi_{\0}$ ($\theta_{\0}<\arcsin\frac{n}{\sqrt{2}}$) the one  at the down dielectric/air interfaces, $\psi_{\0}$ the refraction angle and $\theta_{\0}$ the incidence one
at the left $\overline{AB}$ interface, $\sin\theta_{\0}=n\,\sin\psi_{\0}$.

Due to the fact that the left ($\overline{AB}$) and right ($\overline{CD}$) interfaces have discontinuities along the $\widetilde{z}$-axis (the perpendicular direction to the left/right prism boundaries), it is convenient, as it is used to be done in Quantum Mechanics\cite{AJPBlock}, to introduce, for the transversal block [see Fig.\,\ref{fig:Fig0}(a)] the coordinate system $(x\,,\,\widetilde{y}\,,\,\widetilde{z})$. The same can be done for the up ($\overline{BC}$) and down ($\overline{AD}$) dielectric/air discontinuities whose normal is along the $z_{_{*}}$-axis, by introducing the coordinate system  $(x\,,y_{_{*}}\,,\,z_{_{*}})$. The changes in the wave number components occur in the direction perpendicular to the discontinuity and the reflection and transmission coefficients obtained by a like potential step analysis\cite{AJPBlock}.  These reflection and transmission coefficients have to reproduce the well-known Fresnel coefficients  in the plane wave limit  and contain the information on the geometrical structure of the block in their geometrical phase coming from the position in which the air/dielectric or dielectric/air interface are located. The geometrical optical phase is thus the same for transverse electric (TE) and transverse magnetic (TM) waves.

The transmission  coefficients for the transversal block drawn in Fig.\,\ref{fig:Fig0}(a) are given by
\begin{eqnarray}
\label{TE}
T^{^{[TE]}}(k_{x},k_{y}) &=&
\underbrace{\frac{2\, k_{\widetilde{z}}}{q_{\widetilde{z}} + k_{\widetilde{z}}}}_{\rm left\,\,transmission}\,\,\,\underbrace{\frac{2\, q_{\widetilde{z}}}{q_{\widetilde{z}} + k_{\widetilde{z}}}}_{\rm right\,\,transmission}\,\,\, \underbrace{\left(\,\frac{q_{z_{_{*}}} -\,\, k_{z_{_{*}}}}{q_{z_{_{*}}} +\,\, k_{z_{_{*}}}}\,\right)^{{\2}}}_{\rm up/down\,\,reflections}\nonumber \\
 & = & \frac{4\,k_{\widetilde{z}}\, q_{\widetilde{z}}}{(\,q_{\widetilde{z}} + k_{\widetilde{z}})^{^{\,2}}}\,\left(\,\frac{q_{z_{_{*}}} -\,\, k_{z_{_{*}}}}{q_{z_{_{*}}} +\,\, k_{z_{_{*}}}}\,\right)^{{\2}}\,\,,
\end{eqnarray}
and
\begin{eqnarray}
\label{TM}
T^{^{[TM]}}(k_{x},k_{y}) &=&\frac{4\,\,n^{^{2}}\,k_{\widetilde{z}}\, q_{\widetilde{z}}}{(\,q_{\widetilde{z}} + n^{^{2}}\,k_{\widetilde{z}})^{^{\,2}}}\,\left(\,\frac{q_{z_{_{*}}} -\,\,n^{^{2}} k_{z_{_{*}}}}{q_{z_{_{*}}} +\,\, n^{^{2}}k_{z_{_{*}}}}\,\right)^{{\2}}\,\,,
\end{eqnarray}
where
\begin{eqnarray*}
\left(\,k_{z_{_{*}}},\, k_{\widetilde{y}}\,,\,  k_{\widetilde{z}}\,\right) &=&  \left(\,\sqrt{k^{^{2}} - k_{x}^{^2} - q_{y_{_{*}}}^{^2}}\,,\,k_y\,\cos\theta_{\0} +k_z\,\sin \theta_{\0}\,,\, -\,k_y\,\sin\theta_{\0} +k_z\,\cos\theta_{\0} \,\right)\,\,,\\
\left(\,q_{y_{_{*}}}\,,\,q_{z_{_{*}}}\,,\,q_{\widetilde{z}}\,\right) &=& \left(\,\frac{\,k_{\widetilde{y}}\,+\,q_{\widetilde{z}}}{\sqrt{2}} ,\,\frac{-\,k_{\widetilde{y}}\,+\,q_{\widetilde{z}}}{\sqrt{2}} \,,\,\sqrt{n^{^2}k^{^{2}} - k_{x}^{^2} - k_{\widetilde{y}}^{^2}}\,\right)\,\,.
\end{eqnarray*}
At the center of the wave number distribution, $(k_x,k_y)=(0,0)$, we have
\[k_{\widetilde{z}}=k\,\cos\theta_{\0}\,\,,\,\,\,\,\,
q_{\widetilde{z}}=k\,\sqrt{n^{^{2}}-\sin^{\2}\theta_{\0}}=n\,k\,\cos\psi_{\0}\,\,,\,\,\,\,\,
k_{z_{_{*}}}=k\,\sqrt{1-n^{^{2}} \sin^{\2}\varphi_{\0}}\,\,,\,\,\,\,\,q_{z_{_{*}}}=n\,k\,\cos\varphi_{\0}\,\,,
\]
recovering the well-known Fresnel coefficients\cite{Saleh2007}.

Let us now obtain the geometrical optical phase. When doing it, we have to observe that the continuity equation for the electric field imposes the presence of exponential factors in the reflection and transmission coefficients, taking into account the normal distance between the discontinuities\cite{AJPBlock}. For the reflection coefficient the exponential factor contains the term
\[ 2 \times {\rm reflection\,\, wave\,\, number} \times {\rm normal\,\,distance} = 2\,\,q_{z_{_{*}}}\,\frac{\overline{AB}}{\sqrt{2}} \]
and for the transmission coefficient
\[ {\rm (transmission\,-\,incidence)\,\, wave\,\, number} \times {\rm normal\,\,distance} = (\,q_{\widetilde{z}}-k_{\widetilde{z}}\,)\,\frac{\overline{BC}}{\sqrt{2}}\,\,, \]
for detail see refs.\cite{AJPBlock,2016JMOTSB}. Finally, the geometrical phase for the transversal bock [TB] is given by
\begin{eqnarray}
\Phi_{_{\rm GEO}}^{^{\rm [TB]}}(k_x,k_y) &=& \sqrt{2} \,q_{z_{_{*}}}\, \overline{AB}\,+\, \left(\,q_{\widetilde{z}} - k_{\widetilde{z}}\,\right)\,\frac{\overline{BC}}{\sqrt{2}}  \nonumber \\
& = & \left[\,q_{\widetilde{z}} \,(\,1+\tan\varphi_{\0}\,) - k_{\widetilde{z}}\,\tan\varphi_{\0}-k_{\widetilde{y}}\,\right]\,\overline{AB}\,\,.
\label{block}
\end{eqnarray}
The first order Taylor expansion of this phase is responsible  for the optical path of the beam and reproduces the result predicted by the Snell and reflection laws of Geometric Optics. Indeed, in the integrand of the beam transmitted through a transversal dielectric block now appears the following term
\[ k_x\,x \,+\,k_y\,y -\,\frac{k_x^{^{2}}+k_y^{^{2}}}{2\,k}\,z \,+\, \Phi_{_{\rm GEO}}^{^{\rm [TB]}}(k_x,k_y)  \]
whose first order Taylor expansion in the center of the wave number distribution $(k_x,k_y)=(0,0)$ is
\[ k_x\,\left\{\,x \,+\,\left[\,\frac{\partial \Phi_{_{\rm GEO}}^{^{\rm [TB]}}}{\partial k_x}\right]_{_{(0,0)}}\,\right\} \,+\, k_y\,\left\{\, y \,+\, \left[\,\frac{\partial \Phi_{_{\rm GEO}}^{^{\rm [TB]}}}{\partial k_y}\right]_{_{(0,0)}}\,\right\}
\,\,.
\]
Observing that
\begin{eqnarray*}
\left[\,\frac{\partial \Phi_{_{\rm GEO}}^{^{\rm [TB]}}}{\partial k_x}\right]_{_{(0,0)}} &=&
\left[\,-\,\frac{k_x}{q_{\widetilde{z}}} \,(\,1+\tan\varphi_{\0}\,) \,+\, \frac{k_x}{k_{\widetilde{z}}}\,\tan\varphi_{\0}\,+\,\frac{k_x}{k_z}\,\sin\theta_{0}\,
\right]_{_{(0,0)}}\overline{AB}\,=\,0\,\,,
\end{eqnarray*}
and
\begin{eqnarray*}
\left[\,\frac{\partial \Phi_{_{\rm GEO}}^{^{\rm [TB]}}}{\partial k_x}\right]_{_{(0,0)}}& =&
\left[\,-\,\frac{k_{\widetilde{y}}}{q_{\widetilde{z}}} \,(\,1+\tan\varphi_{\0}\,)\,\cos\theta_{\0} \,+\, \frac{k_{\widetilde{y}}}{k_{\widetilde{z}}}\,\tan\varphi_{\0}\,\cos\theta_{\0}\,-\,\cos\theta_{\0}\,
\right]_{_{(0,0)}}\overline{AB}\\
&=& \left[\,-\,\tan\psi_{\0} \,(\,1+\tan\varphi_{\0}\,)\,\cos\theta_{\0} \,+\, \sin \theta_{\0}\,\tan\varphi_{\0}\,-\,\cos\theta_{\0}\,
\right]\,\,\overline{AB}\\ \\
&=& (\,\sin\theta_{\0}\,-\,\cos\theta_{\0}\,)\,\tan\varphi_{\0}\,\,\overline{AB}\,\,,
\end{eqnarray*}
we obtain that the center of the beam located for the incident beam at $\{x,y\}=\{0,0\}$ is moved for the beam transmitted to a transversal block with $\overline{BC}=\sqrt{2}\,\tan\varphi_{\0}\,\overline{AB}$ at
\[\{\,x\,,\,y\,\}_{_{\rm [TB]}}=\left\{\,0\,,\,  (\,\cos\theta_{\0}\,-\,\sin\theta_{\0}\,)\,\tan\varphi_{\0}\,\,\overline{AB}  \,\right\}\,\,,\]
and this is exactly the prediction of Geometric Optics by using the Snell and reflection laws\cite{AJPBlock}. Is is also interesting to observe that for an incidence greater than the critical one the Fresnel coefficient gains an additional phase
\begin{eqnarray}
\label{TETM}
T^{^{[TE]}}(k_{x},k_{y}) &=&
\frac{4\,q_{\widetilde{z}}\, q_{\widetilde{z}}}{(\,q_{\widetilde{z}} + k_{\widetilde{z}})^{^{\,2}}}\,\exp\left[-\,2\,i\,\frac{|k_{z_{_{*}}}|}{q_{z_{_{*}}}}\,\right]
\,\,,\nonumber \\
T^{^{[TM]}}(k_{x},k_{y}) &=& \frac{4\,\,n^{^{2}}\,k_{\widetilde{z}}\, q_{\widetilde{z}}}{(\,q_{\widetilde{z}} + n^{^{2}}\,k_{\widetilde{z}})^{^{\,2}}}\,\,
\exp\left[-\,2\,i\,n^{^{2}} \frac{|k_{z_{_{*}}}|}{q_{z_{_{*}}}}\,\right]\,\,,
\end{eqnarray}
whose first order Taylor expansion leads to an additional lateral displacement, proportional to $\lambda$, known as Goos-H\"anchen shift\cite{GHS1947,1948AP437}.

It is clear that the second order Taylor expansion of the geometrical  phase will affect the axial behavior ($z$ component) of the optical beam. The study of this contribution for transversal and parallel blocks will be the subject matter of the next section, where, to shorten the calculations, we introduce the angular notation, i.e. we will exchange the cartesian coordinates for the wave number with their spherical counterpart. Consequently the cartesian integration ${\rm d} k_x\,{\rm d}k_y$ will be replaced by the angular integration ${\rm d}\alpha\,{\rm d}\theta$ and the wave number distribution $G(k_x,k_y;\boldsymbol{r})$ by the angular distribution $G(\theta-\theta_{\0},\alpha;\boldsymbol{r})$.

\section*{\color{darkred} \normalsize IV. SECOND ORDER TAYLOR EXPANSION OF THE GEOMETRICAL PHASE}

As anticipated in the previous section, in order to calculate the second order contribution of the geometrical phase (\ref{block}) it is convenient to change the wave number system from cartesian to spherical  coordinates
\begin{equation}
\{\,k_x\,,\,k_y\,,\,k_z\,\,\} = \{\,k\, \sin \alpha\,,\,k\,\sin(\theta-\theta_{\0})\,\cos\alpha\,,\,k\,\cos(\theta-\theta_{\0})\,
\cos\alpha\,\}\,\,.
\end{equation}
By using
\[
\left( \begin{array}{c}
k_{\widetilde{y}} \\
k_{\widetilde{z}}\end{array} \right)  =  \left( \begin{array}{cc}
\cos\theta_{\0} & \sin\theta_{\0} \\
-\sin\theta_{\0} & \cos\theta_{\0} \end{array} \right) \left( \begin{array}{c}
k_y \\
k_z \end{array} \right)\,\,,
\]
we have
\begin{equation}
\{\, k_{\widetilde{y}} \,,\, k_{\widetilde{z}} \,\} = \{\,  k\,\sin \theta\,\cos \alpha \,,\, k \cos \theta \cos \alpha  \,\}\,\,,
\end{equation}
and after simple algebraic manipulation
\begin{equation}
q_{\widetilde{z}} =  n\, k\, \cos \psi \,\cos \alpha \,\,\sqrt{1 + \frac{n^{^2} - 1}{n^{^2}} \left(\frac{\tan \alpha}{\cos \psi}\right)^{\2}} \,\,.
\label{qztil}
\end{equation}
So, in the angular notation, the geometrical optical phase (\ref{block}) is rewritten as
\begin{equation}
\frac{\Phi_{_{\rm GEO}}^{^{\rm [TB]}}(\theta,\alpha)}{k\,\overline{AB}}=  \left\{\,n\, \cos \psi\,\, \sqrt{1 + \frac{n^{^2} - 1}{n^{^2}} \left(\frac{\tan \alpha}{\cos \psi}\right)^{\2}}\, \left(\,1 + \tan \varphi_{\0} \,\right) \,-\, \sin \theta\, -\, \cos \theta\, \tan \varphi_{\0} \,\right\}\,\cos \alpha\,\,.
\end{equation}
Then, the beam transmitted through a transversal dielectric block, see Fig.\,\ref{fig:Fig0}(a), has in its integrand function the  following phase
\begin{equation}
\Psi_{_{\rm TRA}}^{^{\rm [TB]}}(\theta,\alpha,\boldsymbol{r})=
k\,x\,\sin\alpha\,+\,k\,y\,\sin(\theta-\theta_{\0})\,\cos\alpha\,+\,k\,z\,\cos(\theta-\theta_{\0})\,
\cos\alpha\,+\, \Phi_{_{\rm GEO}}^{^{\rm [TB]}}(\theta,\alpha)\,\,.
\end{equation}
 Let us expand this phase up to the second order around $(\theta,\alpha)=(\theta_{\0},0)$,
\begin{eqnarray*}
\Psi_{_{\rm TRA}}^{^{\rm [TB]}}(\theta,\alpha,\boldsymbol{r})\,=\,
k\,x\,\alpha\,+\,k\,y\,(\theta-\theta_{\0})\,+k\,z\,\,
\left[\,1-\frac{\alpha^{^{2}}+\,\,(\theta-\theta_{\0})^{^{2}}}{2}\,\right]\,+\\
\hspace*{-1.1cm}\left\{\,\left[\,n\,\cos \psi_{\0} -n\,\sin\psi_{\0}\,\psi_{\0}^{^{\prime}}\,(\theta-\theta_{\0})-
n\,\left(\sin\psi_{\0}\,\psi_{\0}^{^{\prime\prime}}+\cos\psi_{\0}\,\psi_{\0}^{^{\prime\,2}}\right)
\frac{(\theta-\theta_{\0})^{^{2}}}{2}\,
\right]  \left(1 + \frac{n^{^2} - 1}{2\,n^{^2}\cos^{\2}\psi_{\0}} \frac{\alpha^{^{2}}}{2}\,\right)\, \left(1 + \tan \varphi_{\0}\right) \right.\\
\hspace*{-1.1cm}\left. -\,(\sin\theta_{\0} +\cos\theta_{\0}\tan\varphi_{\0}) \,+\, (\sin\theta_{\0}\tan\varphi_{\0} -\cos\theta_{\0})\,(\theta-\theta_{\0})\,+\,
(\cos\theta_{\0}\tan\varphi_{\0} +\sin\theta_{\0})\,\frac{(\theta-\theta_{\0})^{^{2}}}{2}\, \,\right\}\,
\left(1-\frac{\alpha^{^{2}}}{2}\right)\,k\,\overline{AB}\,\,.
\end{eqnarray*}
Ordering the term in the previous expression, we can rewrite the integrand phase as follows
\begin{equation*}
\Psi_{_{\rm TRA}}^{^{\rm [TB]}}(\theta,\alpha,\boldsymbol{r})= k\,x\,\alpha \,+\, k\,\left[\,y-d(\theta_{\0})\,\overline{AB}\,\right]\,(\theta-\theta_{\0}) \,-\, k\, \left[\,z-f(\theta_{\0})\,\overline{AB}\,\right]\,\frac{\alpha^{^{2}}}{2} \,-\, k\, \left[\,z-g(\theta_{\0})\,\overline{AB}\,\right]\,\frac{(\theta-\theta_{\0})^{^{2}}}{2}\,\,,
\end{equation*}
where
\begin{eqnarray*}
d(\theta_{\0}) &=& n\,\sin\psi_{\0}\,\psi_{\0}^{^{\prime}}\,(1+\tan\varphi_{\0})-\sin\theta_{\0}\,\tan\varphi_{\0}+\cos\theta_{\0}
\,=\, (\cos\theta_{\0}-\sin\theta_{\0})\,\tan\varphi_{\0}\,\,,\\
f(\theta_{\0}) &=& \sin\theta_{\0}+\cos\theta_{\0}\,\tan\varphi_{\0} + \left(\frac{n^{^{2}}-1}{n\,\cos\psi_{\0}} - n\,\cos\psi_{\0}\right)\,(1+\tan\varphi_{\0})\\
 &=&  \sin\theta_{\0}+\cos\theta_{\0}\,\tan\varphi_{\0} - \frac{\cos^{\2}\theta_{\0}}{n\,\cos\psi_{\0}} \,(1+\tan\varphi_{\0})\,\,,
 \\
g(\theta_{\0}) &=& \sin\theta_{\0}+\cos\theta_{\0}\,\tan\varphi_{\0} - n\,\left(\sin\psi_{\0}\,\psi_{\0}^{^{\prime \prime}}+\cos\psi_{\0}\,\psi_{\0}^{^{\prime\,2}}\right)\,(1+\tan\varphi_{\0})\\
&=& (\sin\theta_{\0}+\cos\theta_{\0})\,\tan\varphi_{\0} - \frac{\cos^{\2}\theta_{\0}}{n\,\cos^{\3}\psi_{\0}} \,(1+\tan\varphi_{\0})\,\,.
\end{eqnarray*}
In view of the previous considerations and neglecting the Fresnel phase contribution  which are of the order of $\lambda$, it is possible to give an analytical formula for Hermite-Gaussian beams transmitted through a sequence, $N_y$, of transversal identical blocks and consequently for its intensity.  To shorten our formulas, we introduce the function
\[{\rm HG}_r(a,b)=\frac{\mathrm{w}_{\0}}{\mathrm{w}(b)}\,H^{^{2}}_{r}\left[\frac{\sqrt{2}\,a}{{\rm w}(b)}\right]\,\exp\left[-\,\frac{2\,a^{^{2}}}{{\rm w}(b)}\right]
\]
and assume that the incoming beam is polarized along the $y$ axis.  In this case, the intensity of the transmitted beam is given by
\begin{equation*}
\hspace*{-0.25cm}I_{_{\ell m}}^{^{[\mathrm{NTB}]}} (\boldsymbol{r}) =\left[ \frac{4\,n\,\cos\theta_{\0}\,\cos\psi_{\0}}{(n\,\cos\theta_{\0}+\cos\psi_{\0})^{^{2}}}\right]^{N_y} \,\frac{I_{\0}}{2^{^{\ell+m}}\,\ell!\,m!}\,\,\mathrm{HG}_{\ell}\left(x,z-N_y\,f\,\overline{AB}\right)
\mathrm{HG}_{m}\left(y-N_y\,d\,\overline{AB},z-N_y\,g\,\overline{AB}\right)\,\,.
\end{equation*}
The intensity of the transmitted beam through a sequence, $N_x$, of parallel dielectric blocks, see Fig.\,\ref{fig:Fig0}(b), is easily obtained from the previous one by interchanging $x$ with $y$ and observing that the transmission Fresnel coefficient in this case is the one of the TE waves,
\begin{equation*}
\hspace*{-0.25cm}I_{_{\ell m}}^{^{[\mathrm{NPB}]}} (\boldsymbol{r}) =\left[ \frac{4\,n\,\cos\theta_{\0}\,\cos\psi_{\0}}{(\cos\theta_{\0}+n\,\cos\psi_{\0})^{^{2}}}\right]^{N_x} \,\frac{I_{\0}}{2^{^{\ell+m}}\,\ell!\,m!}\,\,\mathrm{HG}_{\ell}\left(x-N_x\,d\,\overline{AB},z-N_x\,g\,\overline{AB}\right)
\mathrm{HG}_{m}\left(y,z-N_x\,f\,\overline{AB}\right)\,\,.
\end{equation*}
Finally, for a mixed $(N_x,N_y)$ configuration of parallel and transversal blocks, we obtain
\begin{equation}
\label{mixed}
I_{_{\ell m}}^{^{[\mathrm{TRA}]}} (\boldsymbol{r}) = I_{\0}^{^{[\mathrm{TRA}]}}
\,\mathrm{HG}_{\ell}\left[x-d_x(\theta_{\0}),z-z_x(\theta_{\0})\right]\,\,
\mathrm{HG}_{m}\left(y-d_y(\theta_{\0}),z-z_y(\theta_{\0})\right)\,\,,
\end{equation}
where
\[I_{\0}^{^{[\mathrm{TRA}]}} = \left[ \frac{4\,n\,\cos\theta_{\0}\,\cos\psi_{\0}}{(\cos\theta_{\0}+n\,\cos\psi_{\0})^{^{2}}}\right]^{N_x}
\left[ \frac{4\,n\,\cos\theta_{\0}\,\cos\psi_{\0}}{(n\,\cos\theta_{\0}+\cos\psi_{\0})^{^{2}}}\right]^{N_y} \,\frac{I_{\0}}{2^{^{\ell+m}}\,\ell!\,m!}\]
and
\begin{eqnarray*}
\hspace*{-1cm}d_{x}(\theta_{\0})&=& N_{x}\,\,(\cos\theta_{\0}-\sin\theta_{\0})\,\tan\varphi_{\0}\,\overline{AB}\,\,,\nonumber\\
\hspace*{-1cm}d_{y}(\theta_{\0})&=& N_{y}\,\,(\cos\theta_{\0}-\sin\theta_{\0})\,\tan\varphi_{\0}\,\overline{AB}\,\,,\nonumber\\
\hspace*{-1cm}z_x(\theta_{\0})&= & \left[\,(N_x+N_y)\,\cos\theta_{\0}\,\tan\varphi_{\0} + (N_y+N_x\,\tan\varphi_{\0})\,\sin\theta_{\0} - (N_x+N_y\,\cos^{\2}\psi_{\0})\,\frac{\cos^{\2}\theta_{\0}}{n\,\cos^{\3}\psi_{\0}}\,
(1+\tan\varphi_{\0})\,\right]\,\overline{AB}\,\,,\\
\hspace*{-1cm}z_y(\theta_{\0})&= & \left[\,(N_y+N_x)\,\cos\theta_{\0}\,\tan\varphi_{\0} + (N_x+N_y\,\tan\varphi_{\0})\,\sin\theta_{\0} - (N_y+N_x\,\cos^{\2}\psi_{\0})\,\frac{\cos^{\2}\theta_{\0}}{n\,\cos^{\3}\psi_{\0}}\,
(1+\tan\varphi_{\0})\,\right]\,\overline{AB}\,\,.\nonumber
\end{eqnarray*}
The center of the transmitted beam is shifted with respect to the incident one from ($0,0$) to ($d_x,d_y$) and this is in perfect agreement with the prediction of Geometric Optics. What cannot be predicted by the Geometric Optics is obviously the shift observed by the axial coordinate $z$
which affects the beam waist behavior  in the plane $xz$ and $yz$. In the next section, this phenomenon will be analyzed in detail  for a particular incidence angle.

\section*{\color{darkred} \normalsize V. AXIAL BEHAVIOR OF THE TRANSMITTED BEAM}
\begin{figure}
\hspace*{-1.5cm}
\includegraphics[scale=0.9]{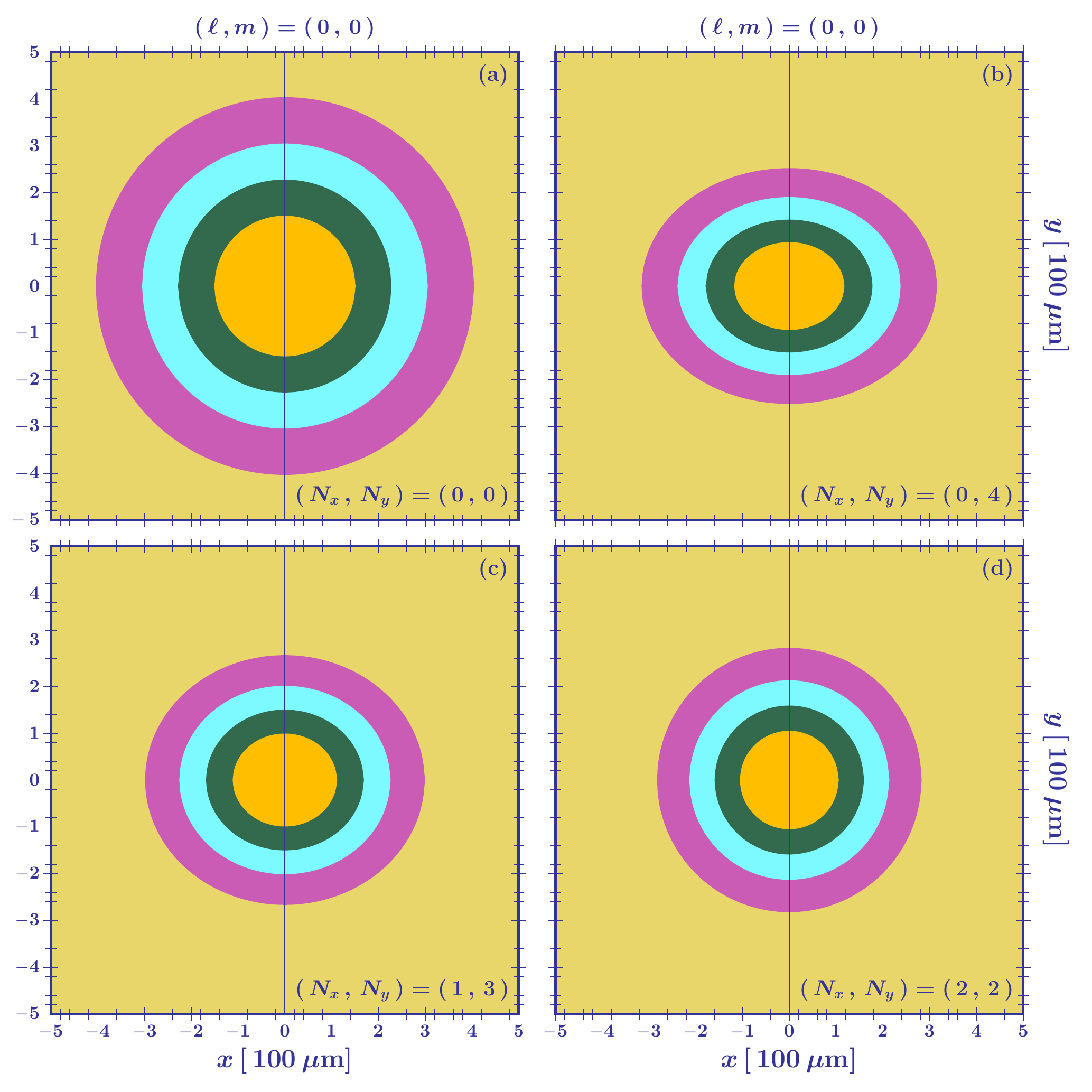}
 \caption{{\bf Gaussian propagation through BK7 materials}. Contour plots $(0.2,\,0.4,\,0.6,\,0.8)$ of the intensity distribution for free (a) and transmitted (b-d) beams. The incidence angle is $\theta=\pi/4$, $\mbox{w}_{\0} =200\,\mu{\rm m}$, $\lambda=0.633\,\mu{\rm}$ and the camera positioned at $z=40\,{\rm cm}$. The
  blocks configuration is $(N_{_{x}}\,,\,N_{_{y}}) = (0,0)$ in (a), $(N_{_{x}}\,,\,N_{_{y}}) = (0,4)$ in (b), $(N_{_{x}}\,,\,N_{_{y}}) = (1,3)$ in (c) and $(N_{_{x}}\,,\,N_{_{y}}) = (2,2)$ in (d). The maximal breaking of symmetry is found in (b) and the delayed axial spreading can be seen in (d) for the configuration $(N_{_{x}}\,,\,N_{_{y}}) = (2,2)$.}
\label{fig:Fig1}
\end{figure}

\begin{figure}
\hspace*{-1.5cm}
\includegraphics[scale=0.9]{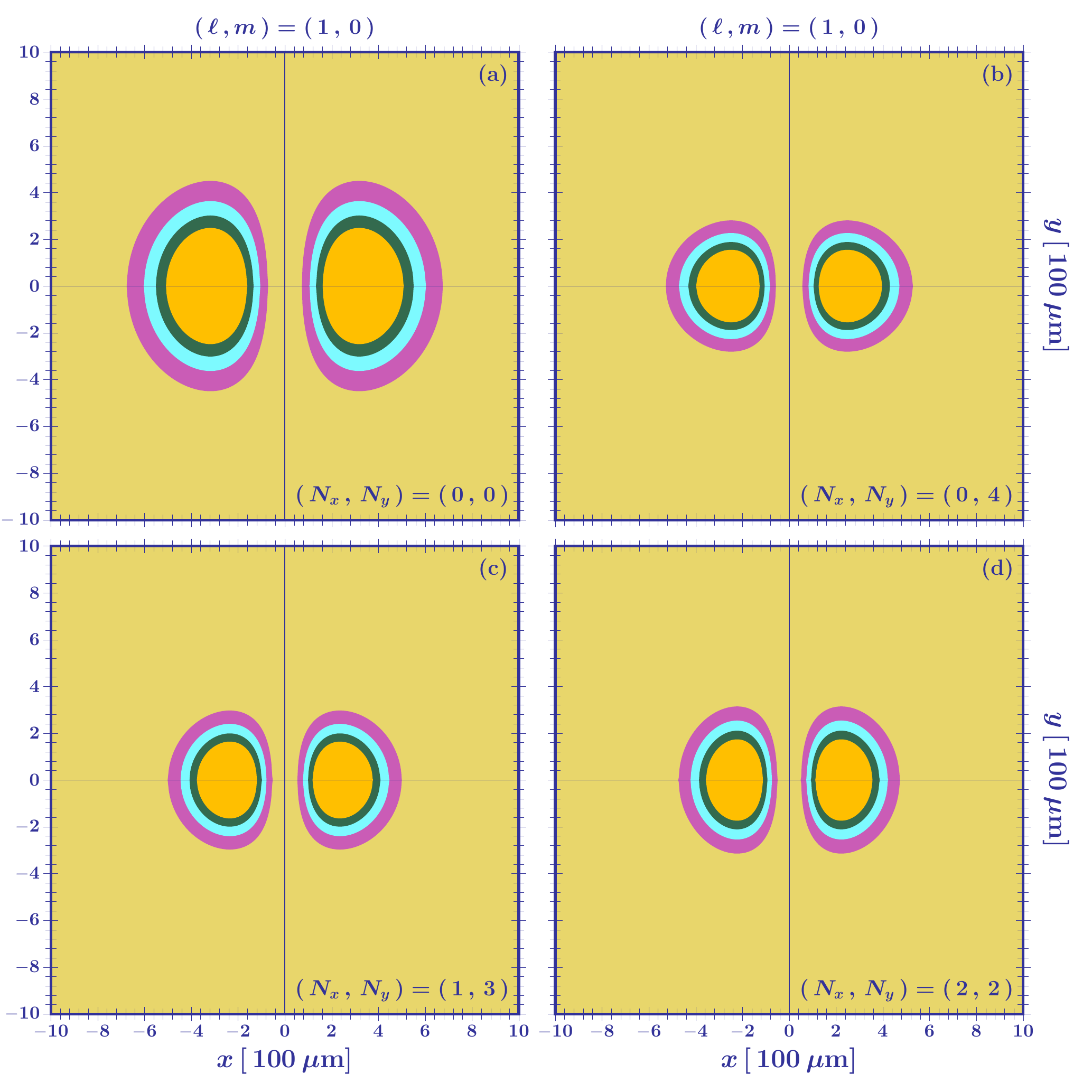}
\caption{{\bf Hermite-Gaussian propagation through BK7 materials}. The same as in Figure \ref{fig:Fig1} for the mode $(1,0)$.}
\label{fig:Fig2}
\end{figure}

\begin{figure}
\hspace*{-1.5cm}
\includegraphics[scale=0.9]{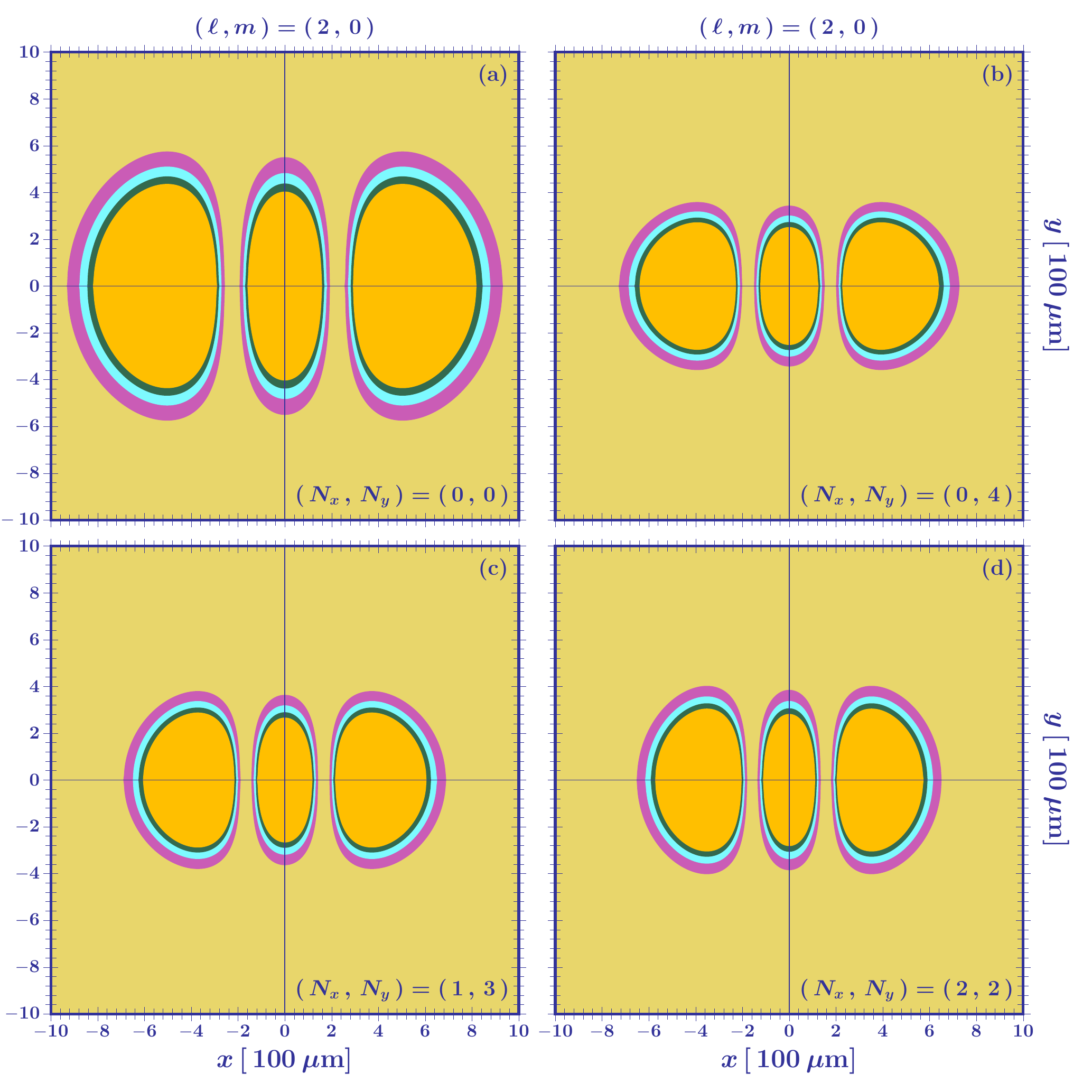}
\caption{{\bf Hermite-Gaussian propagation through BK7 materials}. The same as in Figure \ref{fig:Fig1} for the mode $(2,0)$.}
\label{fig:Fig3}
\end{figure}

Let us consider an incidence angle of $\pi/4$. In this case, for a block with
\[
\overline{BC}\,=\sqrt{2}\,\frac{\sqrt{2\,n^{^{2}}-1}+1}{\sqrt{2\,n^{^{2}}-1}-1}\,\overline{AB}\,\,,
\]
we have
\begin{equation}
\{\,d_x\,,\,d_y\,\}=\{\,0\,,\,0\,\}\,\,,
\end{equation}
and
\begin{equation}
\left\{\,z_x\,,\,z_y\,\right\}  =  \left\{\, N_{_{x}}\,\displaystyle{\frac{2n^{^{2}} + \sqrt{2n^{^{2}} - 1}}{2n^{^{2}} - 1}} + N_{_{y}} \,,\, N_{_{x}} + N_{_{y}}\,\displaystyle{\frac{2n^{^{2}} + \sqrt{2n^{^{2}} - 1}}{2n^{^{2}} - 1}} \, \right\}\,\sqrt{2}\,\,\overline{AB}\,\,.
\label{zxzy}
\end{equation}
For BK7 dielectric blocks, $n=1.515$ (when $\lambda=633\,\mathrm{nm}$), with  $\overline{AB}=2\,\mathrm{cm}$ and $\overline{BC}=9.15\,\mathrm{cm}$, we have
\begin{equation}
\left\{\,z_x\,,\,z_y\,\right\}_{_{\mathrm{BK7}}}  =  \left\{\, 1.806\,N_{_{x}} + N_{_{y}} \,,\, N_{_{x}} + 1.806\,N_{_{y}}\, \right\}\,2\,\sqrt{2}\,\,\mathrm{cm}\,\,.
\label{bk7}
\end{equation}
In order to quantify the axial effects induced on the optical beam by the propagation through different  dielectric blocks configurations, let us, for example, consider the level curve $C$ for a Gaussian beam,
\begin{equation}
\exp\left\{\,- 2 \left[\displaystyle{\frac{x^{^{2}}}{\mbox{w}^{^{2}}(z - z_x)}}\, + \displaystyle{\frac{y^{^{2}}}{\mbox{w}^{^{2}}(z - z_x)}}\right]\right\}=C\,\,.
\end{equation}
For the transmitted beam, the contour plot in the $xy$ plane will be an ellipse,  with semi-axes given by
\begin{equation}
\{\,a_x^{^{[C]}}\,,\,a_y^{^{[C]}}\,\} = \{\,\mbox{w}(z - z_x)\,,\,\mbox{w}(z - z_y) \,\}\,\sqrt{\frac{1}{2}\,\log\frac{1}{C}}\,\,,
\end{equation}
and this ellipse has to be compared with the circle of ray
\[ r^{^{[C]}}= \mbox{w}(z)\,\sqrt{\frac{1}{2}\,\log\frac{1}{C}} \]
which characterizes the free propagation. For an incident beam of waist $\mathrm{w}_{\0}=200\,\mu\mathrm{m}$ and wavelength $\lambda=633\,\mathrm{nm}$ ($z_{_{\mathrm{R}}}=20\,\mathrm{cm}$) and a camera positioned at  $z=40\,{\rm cm}$, we find for the level curve $0.2$ of the free beam the following ray
\begin{equation}
r^{^{[0.2]}}\approx\,\,200\,\,\sqrt{5}\,\,\sqrt{0.8}\,\,\mu {\rm m}\, =\, 400\,\mu{\rm m}\,\,,
\end{equation}
see Fig.\,\ref{fig:Fig1}(a). The beam transmitted through the different blocks configurations of
Fig.\,\ref{fig:Fig1}(b-d)  will suffer, with respect the  free one, the following modifications
\begin{equation}
\{\,a_x^{^{[0.2]}}\,,\,a_y^{^{[0.2]}}\,\}=\left\{
\begin{array}{ll}
\{\,315.3\,,\,251.9\,\}\,\mu{\rm m} & {\rm for}\,\,N_x=0\,\,\,{\rm and}\,\,\,N_y=4\,\,,\\
\{\,299.6\,,\,266.8\,\}\,\mu{\rm m} & {\rm for}\,\,N_x=1\,\,\,{\rm and}\,\,\,N_y=3\,\,,\\
\{\,282.4\,,\,282.4\,\}\,\mu{\rm m} & {\rm for}\,\,N_x=2\,\,\,{\rm and}\,\,\,N_y=2\,\,.
\end{array}
\right.
\end{equation}
The transversal symmetry breaking is thus observed when the blocks configuration is not symmetric in $N_x$ and $N_y$. Increasing  the difference between $N_x$ and $N_y$, we increase the breaking of symmetry. It is also interesting to observe that with an equal number of transversal blocks, $N_x=N_y$, we recover the symmetry and we delay the axial spreading of the beam, see Fig.\,\ref{fig:Fig1}(d). In Figs.\ref{fig:Fig2}-\ref{fig:Fig3}, we plot the free and transmitted Hermite-Gaussian beam for higher modes.

\section*{\color{darkred} \normalsize VI. CONCLUSIONS}

In this article, we have studied the transversal breaking of symmetry and the axial spreading modification due to the second order contribution of the geometrical part of the optical  phase for Hermite Gaussian beams. Based on the second Taylor expansion, we showed the possibility to get an analytical expression for the transmitted beam intensity. The choice of $\theta=\pi/4$ was done in view of a possible experimental implementation. This incidence angle also allows to compare in an easy way the free to the transmitted beam. The analysis presented in this paper for a sequence of transversal and parallel dielectric blocks represents only a first step towards the understanding of the intriguing phenomenon of the breaking of symmetry and surely deserves further studies, in particular in proximity of the critical incidence region were analytical approximation needs to be done in the appropriate way.

\vspace*{0.5cm}

\noindent
\textbf{\color{darkred} \normalsize ACKNOWLEDGEMENTS}\\
The authors thank M. P. Araujo and G. G. Maia for interesting comments and stimulating discussions which motivated the study of the effect of the optical phase on propagation of Hermite-Gaussian beams in dielectric medium.  The authors also thank  Fapesp, CNPq, and Faperj for the financial support.

\newpage

\newpage

\newpage

\newpage

\end{document}